\documentclass[twocolumn,prb,showpacs,preprintnumbers,amsmath,amssymb]{revtex4}
\usepackage{graphicx}
\usepackage{mathptmx,bm}

\begin{document}


\title{Absence of Superconductivity in the ``hole-doped" Fe pnictide Ba(Fe$_{1-x}$Mn$_{x}$)$_{2}$As$_{2}$: Photoemission and X-ray Absorption Spectroscopy Studies}

\author{H. Suzuki$^{1}$, T. Yoshida$^{1,2}$, S. Ideta$^1$, G. Shibata$^{1}$, K. Ishigami$^{1}$, T. Kadono$^{1}$, A. Fujimori$^{1,2}$, M. Hashimoto$^3$, D. H. Lu$^3$, Z.-X. Shen$^3$,
 K. Ono$^4$, E. Sakai$^4$, H. Kumigashira$^4$, M. Matsuo$^{5}$ and T. Sasagawa$^{5}$}
 
\affiliation{$^1$Department of Physics, University of Tokyo,
Bunkyo-ku, Tokyo 113-0033, Japan}

\affiliation{$^2$JST, Transformative Research-Project on Iron
Pnictides (TRIP), Chiyoda, Tokyo 102-0075, Japan}

\affiliation{$^3$Stanford Synchrotron Radiation Lightsource, Stanford University, Stanford, CA 94305, USA}

\affiliation{$^4$KEK, Photon Factory, Tsukuba, Ibaraki 305-0801, Japan}

\affiliation{$^5$Materials and Structures Laboratory, Tokyo institute of Technology, Kanagawa 226-8503, Japan}

\date{\today}

\begin{abstract}
We have studied the electronic structure of Ba(Fe$_{1-x}$Mn$_{x}$)$_{2}$As$_{2}$ ($x$=0.08), which fails to become a superconductor in spite of the formal hole doping like Ba$_{1-x}$K$_{x}$Fe$_{2}$As$_{2}$, by photoemission spectroscopy and X-ray absorption spectroscopy (XAS). With decreasing temperature, a transition from the paramagnetic phase to the antiferromagnetic phase was clearly observed by angle-resolved photoemission spectroscopy. XAS results indicated that the substituted Mn atoms form a strongly hybridized ground state. Resonance-photoemission spectra at the Mn $L_{3}$ edge revealed that the Mn $3d$ partial density of states is distributed over a wide energy range of  2-13 eV below the Fermi level ($E_F$), with little contribution around $E_F$. This indicates that the dopant Mn 3$d$ states are localized in spite of the strong Mn $3d$-As $4p$ hybridization and split into the occupied and unoccupied parts due to the on-site Coulomb and exchange interaction. The absence of superconductivity in Ba(Fe$_{1-x}$Mn$_{x}$)$_{2}$As$_{2}$ can thus be ascribed both to the absence of carrier doping in the FeAs plane, and to the strong stabilizaiton of the antiferromagnetic order by the Mn impurities.
\end{abstract}


\maketitle

\begin{figure}[htbp]
\begin{center}
\includegraphics[width=6cm]{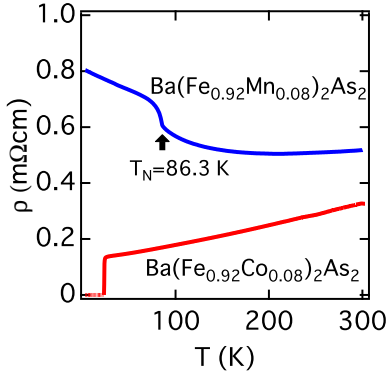}
\caption{(Color online) Resisivity curves of Ba(Fe$_{0.92}$Mn$_{0.08}$)$_{2}$As$_{2}$ (Mn-Ba122) and Ba(Fe$_{0.92}$Co$_{0.08}$)$_{2}$As$_{2}$. }
\label{resis}
\end{center}
\end{figure}

 
Since the discovery of high-$T_{c}$ superconductivity in the iron pnictides \cite{Kamihara.Y_etal.Journal-of-the-American-Chemical-Society2008}, various types of carrier doping have been successfully performed to induce superconductivity. Superconductivity appears in BaFe$_{2}$As$_{2}$  (Ba122) by hole doping realized by K substitution for the Ba sites \cite{Rotter.M_etal.Phys.-Rev.-Lett.2008} or by electron doping realized by Co \cite{Sefat.A_etal.Phys.-Rev.-Lett.2008}, Ni, or Cu substitution \cite{Ni.N_etal.Phys.-Rev.-B2010} for the Fe sites. However, Mn substitution for the Fe sites does not induce superconductivity and antiferromagnetic order persists in Ba(Fe$_{1-x}$Mn$_{x}$)$_{2}$As$_{2}$ (Mn-Ba122) in the entire doping range $x$ below $\sim$ 60 K \cite{Kim.M_etal.Phys.-Rev.-B2010, Thaler.A_etal.Phys.-Rev.-B2011}. The orthorhombic distortion of Ba122 disappears at $x\sim0.09$ and the system enters a tetragonal phase. As shown in Fig. \ref{resis}, the resistivity of Mn-Ba122 \cite{Kim.M_etal.Phys.-Rev.-B2010} shows a jump at the magneto-structural phase transition temperature and gradually increases below it. This behavior is contrasted with the metallic behaviors of the parent compound and the hole-doped Ba$_{1-x}$K$_{x}$Fe$_{2}$As$_{2}$ (K-Ba122) \cite{Ohgushi.K_etal.Phys.-Rev.-B2012}, where the resistivity drops below the transition temperature. A neutron scattering experiment on Mn-Ba122 \cite{Tucker.G_etal.Phys.-Rev.-B2012} has revealed competition between the stripe-type spin-density-wave (SDW) order [$\bm{Q}_{\text{stripe}}$=($\frac{1}{2},\frac{1}{2},1$)] seen in the parent compound and a G-type antiferromagnetic order [$\bm{Q}_{\text{G-type}}$=($1,0,1$)] predominant in the tetragonal phase; in the orthorhombic phase, there are additional inelastic spin excitation peaks which represent the G-type antiferromagnetic fluctuations. This result indicates that Mn doping predominantly introduces an additional local magnetic order which is distinct from the magnetic order of the parent compound, instead of suppressing the long-range SDW order of the parent compound. It is theoretically suggested that G-type magnetic fluctuations strongly suppress the $s^{\pm}$ superconducting state \cite{Fernandes.R_etal.Phys.-Rev.-Lett.2013}.
According to the proximity scenario of iron-based superconductors to Mott insulators \cite{Misawa.T_etal.Phys.-Rev.-Lett.2012,Ishida.H_etal.Phys.-Rev.-B2010}, Mn doping into the Fe sites is thought to drive the system towards a Mott insulator with the $d^{5}$ electronic configuration. In fact, the end member BaMn$_{2}$As$_{2}$ is a G-type antiferromagnetic insulator with an ordered magnetic moment of $\mu = 3.88 \mu_{\text{B}}$/Mn at 10 K directed along the $c$ axis \cite{Singh.Y_etal.Phys.-Rev.-B2009,Johnston.D_etal.Phys.-Rev.-B2011}. Thus the clarification of the electronic structure associated with the two distinct magnetic orderings may give useful insight into the doping mechanism and superconductivity in the 122 systems.

In this study, we have studied the electronic structure of Mn-Ba122 ($x$=0.08) with angle-resolved photoemission spectroscopy (ARPES) and X-ray absorption spectroscopy (XAS). ARPES spectra above and below the transition temperature visualize the folding of the band structure due to the stripe-type antiferromagnetic order. Comparison of the XAS spectra at the Mn $2p$-$3d$ absorption edge with those of other reference compounds has made clear that XAS spectra at the Mn and Fe $L_{2,3}$ edges (2$p$ to 3$d$ absorption edges) have similar line shapes characteristic of metallic or strongly $p$-$d$ hybridized compounds. The partial densities of states of Fe 3$d$ and Mn 3$d$ have been obtained by resonance photoemission spectroscopy (RPES) using photon energies at the Mn and Fe $L_{3}$ absorption edges. Based on the obtained spectroscopic data, we shall discuss the origin of the absence of superconductivity in Mn-Ba122 and the nature of electron correlation in the Fe-based superconductors.

Ba(Fe$_{0.92}$Mn$_{0.08}$)$_{2}$As$_{2}$ single crystals were grown by the self-flux technique in evacuated double quartz tubes \cite{Nakashima.Y_etal.Physica-C:-Superconductivity2010}. Thin plate crystals having $ab$-surfaces of the $\sim 2 \times 2$ mm were separated from the flux by using the cleavage property of the material along the $ab$-planes. 
Resistivity along the $ab$-plane was measured by the four-probe method using a home-built apparatus with a closed circle He-refrigerator. The resistivity curves of Mn-Ba122 ($x$=0.08) and Ba(Fe$_{0.92}$Co$_{0.08}$)$_{2}$As$_{2}$ (Co-Ba122) are compared in Fig. \ref{resis}. The N\'eel temperature $T_{N}=86.3$ K is determined as the temperature at which the resistivity shows a discontinuous jump. The normal-state resistivity of Mn-Ba122 is about twice larger than that of Co-Ba122, indicative of stronger impurity potential of the Mn atom than that of the Co atom.

ARPES experiments were performed at Beamline 5-4 of Stanford Synchrotron Radiation Lightsource using an R4000 Scienta electron analyzer. ARPES spectra were recorded using 31 eV photons under a pressure better than $1\times 10^{-10}$ Torr. The total energy resolution was set to 12 meV or better and the angular resolution was 0.3 degrees. XAS and RPES experiments were performed at Beamilne 2-C of Photon Factory, High-Energy Accelerator Research Organization (KEK). Mn-Ba122 single crystals were cleaved \textit{in situ} to obtain fresh surfaces in all the measurements.

\begin{figure}[htbp]
\begin{center}
\includegraphics[width=8cm]{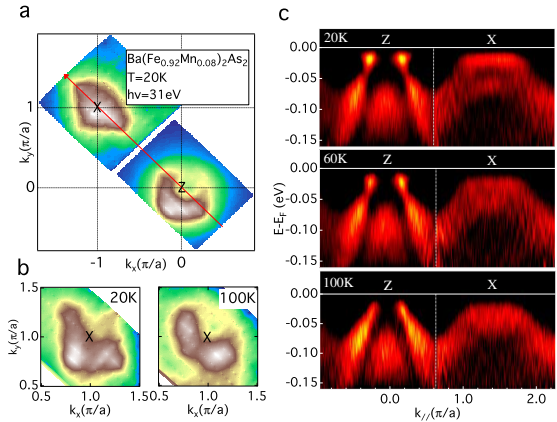}
\caption{(Color online) ARPES spectra of Mn-Ba122. (a) Fermi surface mapping of Mn-Ba122 taken at $T$ = 20 K with 31 eV photons. (b) Fermi surface reconstruction due to antiferromagnetic ordering. (c) Second derivative plot of EDC's taken along the cut from the Z point to the X point shown by a red diagonal line in panel (a).}
\label{mapping}
\end{center}
\end{figure}

Fermi-surface mapping in the $k_{x}$-$k_{y}$ plane is shown in Fig. \ref{mapping}(a). The in-plane lattice constant $a=4.17$ $\text{\AA}$ is used. The $k_{z}$ position ($\sim$ $\frac{7\pi}{c}$) is close to that of the Z point. In order to see the effect of the magneto-structural transition we have performed measurements below and above $T_{N}$=86.3 K. Figure \ref{mapping}(b) shows the reconstruction of the electron-like Fermi surfaces (FSs) around the zone corner (X point) due to the antiferromagnetic ordering: the circular shape of the FSs  above $T_{N}$ (right panel) changes into a propeller-like one below $T_{N}$ (left panel). The broad ARPES intensity distribution in momentum space and the complexity arising from the folding of multiple bands make it difficult to identify the $k_{F}$ positions of each FS. We shall argue below that this broadness originates partly from the strong scattering of quasi-particles by  randomly distributed Mn local spins. Figure \ref{mapping}(c) shows the second derivative plots of the ARPES intensities along the cut shown in panel (a) at three different temperatures. As one decreases the temperature from 100 K to 20 K, the Dirac-cone-like features below $E_{F}$ show up as bright spots near the Z point as a result of band folding and the bands at the X point start to split due to the orthorhombic structural distortion.

\begin{figure}[htbp]
\begin{center}
\includegraphics[width=8cm]{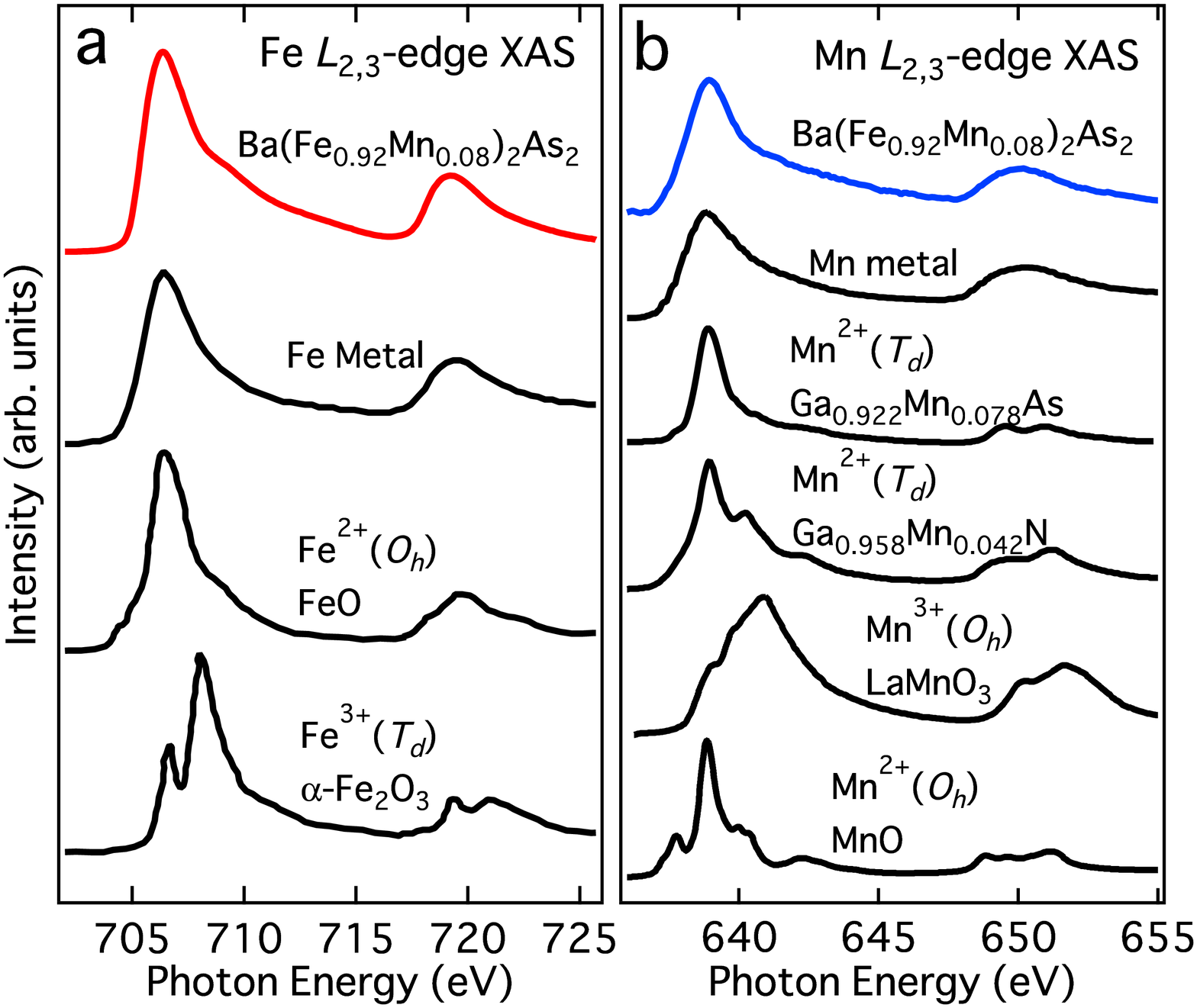}
\caption{(Color online) XAS spectra of Mn-Ba122. (a) Fe $L_{2,3}$-edge XAS spectra of Mn-Ba122 compared with those of Fe metal, FeO, and $\alpha$-Fe$_{2}$O$_{3}$ \cite{Regan.T_etal.Phys.-Rev.-B2001}. (b) Mn $L_{2,3}$-edge XAS spectra of Mn-Ba122 compared with those of Mn metal \cite{Andrieu.S_etal.Phys.-Rev.-B1998}, Ga$_{0.922}$Mn$_{0.078}$As \cite{Takeda.Y_etal.Phys.-Rev.-Lett.2008}, Ga$_{0.958}$Mn$_{0.042}$N \cite{Hwang.J_etal.Applied-Physics-Letters2007}, LaMnO$_{3}$, and MnO \cite{Burnus.T_etal.Phys.-Rev.-B2008}.}
\label{XAS}
\end{center}
\end{figure}

In order to clarify the electronic states of the doped Mn, we have performed XAS measurement in the photon energy regions in the $L_{2,3}$ edges of Mn and Fe. Figure \ref{XAS} shows the Mn and Fe $L_{2,3}$ absorption edges of Mn-Ba122 compared with some reference systems. The peak positions of the Fe $L_{2,3}$-edge XAS spectrum are close to those of FeO, and its line shape is intermediate between those of FeO and Fe metal. These results confirm that Fe atoms in the parent compound has the valence of 2+ in the antiferromagnetic metallic ground state as is widely believed.
The line shapes of the Mn $L_{2,3}$-edge XAS spectrum of Ba(Fe$_{0.92}$Mn$_{0.08}$)$_{2}$As$_{2}$ is intermediate between those of Mn metal and Ga$_{0.922}$Mn$_{0.078}$As (GaMnAs), and does not show fine multiplet structures seen in MnO and Ga$_{0.958}$Mn$_{0.042}$N (GaMnN). The peak position ($\sim$639 eV) is closer to those of MnO, GaMnAs, and GaMnN. The $L_{3}$-edge spectrum of LaMnO$_{3}$ has its weight on higher photon energies ($\sim$642 eV) than that of Mn-Ba122, and has three shoulder structures. These results indicate that the doped Mn-atoms in Mn-Ba122 take the Mn$^{2+}$ state but that the Mn $3d$ electrons are more strongly hybridized with As 4$p$ orbitals in Mn-Ba122 than in GaMnAs.

\begin{figure}[htbp]
\begin{center}
\includegraphics[width=9cm]{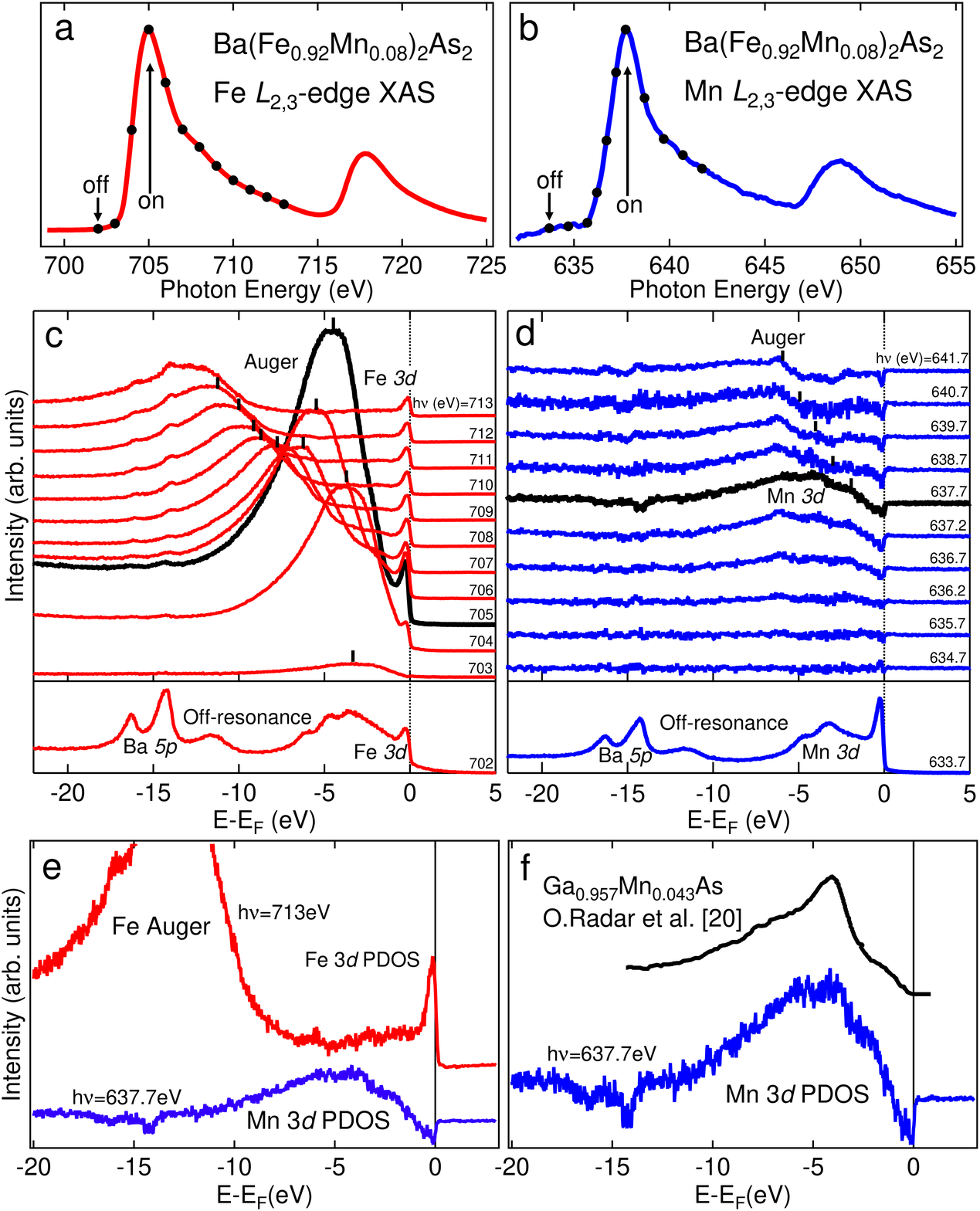}
\caption{(Color online) Resonance photoemission (RPES) spectra of Mn-Ba122. (a), (b) Fe and Mn $L_{2,3}$-edge XAS spectra. Black circles indicate the photon energies used for RPES. On- and off-resonance photon energies are indicated by vertical arrows. (c), (d) Evolution of the Mn-Ba122 valence-band photoemission spectra with photon energy indicated in (a) and (b). Off-resonance photoemission signals shown at the bottom of panels (c), (d) have been subtracted from the original spectra in order to highlight the resonant enhancement of the spectral weight. On-resonance spectra are shown by black curves. Vertical bars indicate constant kinetic energies characteristic of Auger-electron emission. (e) Fe and Mn 3$d$ partial density of states (PDOS) deduced by subtracting the off-resonance spectra from the on-resonance spectra. (f) Mn 3$d$ PDOS of Ga$_{0.957}$Mn$_{0.043}$As \cite{Rader.O_etal.Phys.-Rev.-B2004} compared with that of Mn-Ba122.}
\label{RPES}  
\end{center}
\end{figure}

In resonance photoemission spectroscopy, one makes use of the property that the cross-section of photoemission from an atomic orbital is enhanced by quantum-mechanical interference between direct photoemission of a $d$ electron, $3d^{n}+h\nu\rightarrow 3d^{n-1}+e^{-}$, and absorption followed by a Koster-Kr$\ddot{\text{o}}$nig transition $2p^{6}3d^{n}+h\nu\rightarrow 2p^{5}3d^{n+1}\rightarrow 2p^{6}3d^{n-1}+e^{-}$ \cite{Gelmukhanov.F_etal.Physics-Reports1999,Bruhwiler.P_etal.Rev.-Mod.-Phys.2002}. This effect can be utilized to extract the $3d$ partial density of states of a transition element in solids. Following the observed x-ray absorption spectra, we measured the valence-band photoemission spectra with photon energies from the off-resonance to on-resonance regions. Figure \ref{RPES} shows photoemission spectra taken at photon energies around the Mn and Fe $L_{2,3}$ absorption edges. A series of photoemission spectra taken with a small photon energy interval enable us to identify the Auger-electron emission peak, which have a constant kinetic energy, as well as the enhancement of photoemission features. 
Figures \ref{RPES}(c) and (d) show the valence-band spectra taken with photon energies in the Fe and Mn $L_{2,3}$ absorption regions, respectively. 
The strongest enhancement around the Fe $L_{2,3}$ resonance photon energy mainly originates from the Auger emission process, and its position relative to $E_{F}$ shifts with the incident photon energy as shown by vertical bars. The predominance of the Auger feature at the Fe $L_{3}$ edge arises from the fact that the Fe 3$d$ 
electrons in the FeAs plane have a high DOS at $E_{F}$ and that they screen the core-hole potential immediately after the photo-absorption. On the contrary, the weaker Auger feature 
at the Mn $L_{3}$ edge implies that the Mn 3$d$ electrons are relatively immobile (due to the gap around the $E_{F}$) and that the core hole is not efficiently screened before the Koster-Kr$\ddot{\text{o}}$nig transition occurs. This is in contrast with the strong Auger feature observed 
at the Co $L_{2,3}$ edges of Ca(Fe$_{0.944}$Co$_{0.056}$)$_{2}$As$_{2}$ \cite{Levy.G_etal.Phys.-Rev.-Lett.2012}, which clearly demonstrated the metallic nature of 
Co 3$d$ electrons and the high Co 3$d$ PDOS at $E_{F}$. 

Figure \ref{RPES}(e) shows the deduced Fe and Mn 3$d$ PDOS. While the main feature of the Fe 3$d$ PDOS is located 
within 2 eV from $E_{F}$, the Mn 3$d$ PDOS is widely distributed in the range from $\sim$ 2 eV to $\sim$13 eV below $E_{F}$ with little weight around $E_{F}$. This Mn $3d$ PDOS resembles that of Ga$_{0.957}$Mn$_{0.043}$As as shown in Fig. \ref{RPES}(f) and is another evidence for the strong hybridization between the Mn 3$d$ and As 4$p$ orbitals  and the strong Coulomb and exchange interaction between the Mn $3d$ electrons. The distinct Fe and Mn PDOSs are in striking contrast to 
the strong similarity between the Fe 3$d$ and Co 3$d$ PDOSs found in Ca(Fe$_{0.944}$Co$_{0.056}$)$_{2}$As$_{2}$ \cite{Levy.G_etal.Phys.-Rev.-Lett.2012}, where the center of 
gravity for Co 3$d$ is only $\sim$ 0.25 eV deeper and the spectral line shapes of the Fe and Co PDOS are almost identical. A recent ARPES study of Co, Ni, and Cu-substituted BaFe$_{2}$As$_{2}$ has also shown that the deviation from the rigid-band model is minimal for Co-substitution \cite{Ideta.S_etal.Phys.-Rev.-Lett.2013}.
According to a first-principle calculation on the 
effects of transition-metal substitution in LaFeAsO \cite{Konbu.S_etal.Solid-State-Communications2012}, the on-site potential differences between the impurity 3$d$ and the Fe 3$d$ orbitals of La(Fe,$M$)AsO ($M$: transition metal) shifts monotonically as $M$ goes from Mn, Co, Ni, to Zn ($\sim$ 0.29 for Mn, $\sim$ -0.36 for Co, $\sim$ -0.90 for Ni, and $\sim$ -8.0 for Zn in units of eV). In order to reconcile the present Mn PDOS position $\sim$ 2-13 eV below $E_{F}$ with the calculated positive Mn 3$d$ position, we consider that Hund's coupling and on-site Hubbard interaction between Mn 3$d$ electrons are important and play an essential role. For the $d^{5}$ configuration of the Mn$^{2+}$ ion, the Hubbard splitting into the $\uparrow$ and $\downarrow$ bands is given by $U+4J$ (where $U$ and $J$ are the on-site Coulomb and exchange energy), which amounts to at least $\sim$ 5 eV. Therefore, it is natural for the occupied Mn 3$d$ levels to sink several below $E_{F}$. We, therefore, conclude that a large magnetic moment of $S=5/2$ is formed at the Mn site in Mn-Ba122 and that the magnetism of the Mn $3d$ electrons cannot be described by the weak coupling SDW picture. The large Mn magnetic moment serves to maintain the magnetic ordering of the parent compound up to high Mn concentration. 

\begin{figure}[htbp]
\begin{center}
\includegraphics[width=7cm]{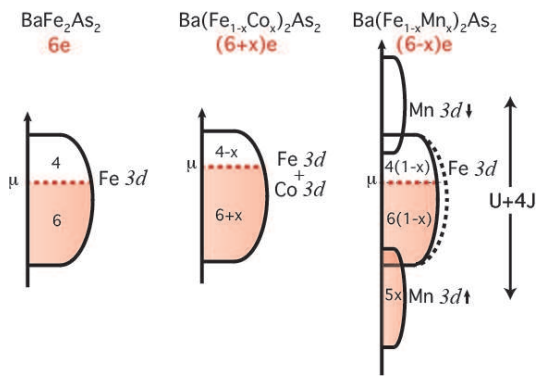}
\caption{(Color online) Schematic density of states of BaFe$_{2}$As$_{2}$, Ba(Fe$_{1-x}$Co$_{x}$)$_{2}$As$_{2}$, and Ba(Fe$_{1-x}$Mn$_{x}$)$_{2}$As$_{2}$. The number of electrons in the $d$ band per unit formula is given for each compound. Co doping introduces additional electrons into the band near $E_{F}$ because the Co 3$d$ and Fe 3$d$ orbitals form a common (Fe+Co) 3$d$ band and the additional $x$ electrons occupy the latter band. On the other hand, Mn doping does not introduce holes into the Fe 3$d$ orbitals if the Mn $3d$ level is split into the occupied majority-spin and unoccupied minority-spin bands due to the Coulomb and exchange interaction. Because the lower Mn 3$d$ level can accommodate 5$x$ electrons, and the DOS of the Fe $3d$ band is reduced by a factor of $1-x$, one finds that the Fermi level does not shift upon Mn substitution. The Mn atoms with the $d^{5}$ configuration has the spin of $S=5/2$. }
\label{DOS}
\end{center}
\end{figure}

In order to explain the absence of itinerant hole carriers in Mn-Ba122, we show in Fig. \ref{DOS} the schematic density of states of BaFe$_{2}$As$_{2}$, Ba(Fe$_{1-x}$Co$_{x}$)$_{2}$As$_{2}$, and Ba(Fe$_{1-x}$Mn$_{x}$)$_{2}$As$_{2}$. Here, the crystal-field splitting and the hybridization with the As 4$p$ orbitals are neglected for simplicity. In the parent compound, six out of the ten Fe 3$d$ orbitals are occupied. In the Co-substituted compound, the Fe 3d and Co 3d orbitals form a common energy bands as a consequence of weak impurity potential of the Co 3$d$ states. Therefore, Co doping actually introduces electrons into the combined Fe 3$d$-Co 3$d$ bands. On the other hand, due to the splitting of the Mn 3$d$ level into the $\uparrow$ and $\downarrow$ levels caused by the on-site Coulomb repulsion and  Hund's coupling, the Fe 3$d$ and Mn 3$d$ orbitals do not form common energy bands. Because the lower Mn $3d$ level accommodate $5x$ electrons and the DOS of the Fe 3d band is reduced by a factor of $1-x$, the Fermi level remains unshifted by the Mn substitution. The absence of carrier doping by transtition-metal substitution is also found in Zn-substitution, where the deep Zn $3d$ orbitals do not form a combined band with the Fe 3$d$ orbitals \cite{Ideta.S_etal.Phys.-Rev.-B2013}.

The emergence of the large local magnetic moments at the Mn sites may prompt us to consider that the upturn behaviour of  Mn-Ba122 resistivity seen at $\sim$ 200 K for $x=0.08$, or at lower temperatures for smaller $x$, is due to the Kondo effect \cite{Urata.T_etal.ArXiv-eprints2013}. However, as suggested by a dynamical mean-field calculation study \cite{Georges.A_etal.Annual-Review-of-Condensed-Matter-Physics2013}, large $J$ in multiband systems significantly reduce the Kondo temperature. Therefore this behaviour is better ascribed to the weak-localization effect caused by the strong random potentials induced by magnetic Mn atoms. 

In conclusion, by using photoemission and X-ray absorption spectroscopy, we have probed the local electronic structure of the Mn dopant in Ba(Fe$_{0.92}$Mn$_{0.08}$)$_{2}$As$_{2}$. The Mn 3$d$ orbitals are strongly hybridized with As 4$p$ orbitals, and the Mn 3$d$ PDOS is widely distributed in the range from $\sim$ 2 eV to $\sim$ 13 eV below $E_{F}$ owing to the energy level splitting by on-site Coulomb interaction and Hund's rule coupling. The energy gap opens at $E_{F}$ in the Mn $3d$ PDOS and hole doping does not occur by Mn substitution.

We are grateful to S. Uchida for enlightening discussions.
This work was supported by the Japan-China-Korea A3 Foresight Program from the
Japan Society for the Promotion of Science  and a Grant-in-Aid for Scientific Research on Innovative
Area ``Materials Design through Computics''. ARPES experiments were performed at the
Stanford Synchrotron Radiation Lightsource, operated by the Office of Basic Energy Science,
US Department of Energy. Experiment at Photon Factory was approved by the Photon Factory Program Advisory
Committee (Proposals No. 2009S2-005 and No. 2011G582).

\bibliography{MnBib}

\begin{thebibliography}{28}%
\makeatletter
\providecommand \@ifxundefined [1]{%
 \@ifx{#1\undefined}
}%
\providecommand \@ifnum [1]{%
 \ifnum #1\expandafter \@firstoftwo
 \else \expandafter \@secondoftwo
 \fi
}%
\providecommand \@ifx [1]{%
 \ifx #1\expandafter \@firstoftwo
 \else \expandafter \@secondoftwo
 \fi
}%
\providecommand \natexlab [1]{#1}%
\providecommand \enquote  [1]{``#1''}%
\providecommand \bibnamefont  [1]{#1}%
\providecommand \bibfnamefont [1]{#1}%
\providecommand \citenamefont [1]{#1}%
\providecommand \href@noop [0]{\@secondoftwo}%
\providecommand \href [0]{\begingroup \@sanitize@url \@href}%
\providecommand \@href[1]{\@@startlink{#1}\@@href}%
\providecommand \@@href[1]{\endgroup#1\@@endlink}%
\providecommand \@sanitize@url [0]{\catcode `\\12\catcode `\$12\catcode
  `\&12\catcode `\#12\catcode `\^12\catcode `\_12\catcode `\%12\relax}%
\providecommand \@@startlink[1]{}%
\providecommand \@@endlink[0]{}%
\providecommand \url  [0]{\begingroup\@sanitize@url \@url }%
\providecommand \@url [1]{\endgroup\@href {#1}{\urlprefix }}%
\providecommand \urlprefix  [0]{URL }%
\providecommand \Eprint [0]{\href }%
\@ifxundefined \urlstyle {%
  \providecommand \doi  [0]{\begingroup \@sanitize@url \@doi}%
  \providecommand \@doi [1]{\endgroup \@@startlink {\doibase
  #1}doi:\discretionary {}{}{}#1\@@endlink }%
}{%
  \providecommand \doi  [0]{doi:\discretionary{}{}{}\begingroup
  \urlstyle{rm}\Url }%
}%
\providecommand \doibase [0]{http://dx.doi.org/}%
\providecommand \Doi [0]{\begingroup \@sanitize@url \@Doi }%
\providecommand \@Doi  [1]{\endgroup\@@startlink{\doibase#1}\@@Doi}%
\providecommand \@@Doi [1]{#1\@@endlink}%
\providecommand \selectlanguage [0]{\@gobble}%
\providecommand \bibinfo  [0]{\@secondoftwo}%
\providecommand \bibfield  [0]{\@secondoftwo}%
\providecommand \translation [1]{[#1]}%
\providecommand \BibitemOpen [0]{}%
\providecommand \bibitemStop [0]{}%
\providecommand \bibitemNoStop [0]{.\EOS\space}%
\providecommand \EOS [0]{\spacefactor3000\relax}%
\providecommand \BibitemShut  [1]{\csname bibitem#1\endcsname}%
\bibitem [{\citenamefont {Kamihara}\ \emph {et~al.}(2008)\citenamefont
  {Kamihara}, \citenamefont {Watanabe}, \citenamefont {Hirano},\ and\
  \citenamefont
  {Hosono}}]{Kamihara.Y_etal.Journal-of-the-American-Chemical-Society2008}%
  \BibitemOpen
  \bibfield  {author} {\bibinfo {author} {\bibfnamefont {Y.}~\bibnamefont
  {Kamihara}}, \bibinfo {author} {\bibfnamefont {T.}~\bibnamefont {Watanabe}},
  \bibinfo {author} {\bibfnamefont {M.}~\bibnamefont {Hirano}}, \ and\ \bibinfo
  {author} {\bibfnamefont {H.}~\bibnamefont {Hosono}},\ }\Doi
  {10.1021/ja800073m} {\bibfield  {journal} {\bibinfo  {journal} {J. Am. Chem.
  Soc.},\ }\textbf {\bibinfo {volume} {130}},\ \bibinfo {pages} {3296}
  (\bibinfo {year} {2008})}\BibitemShut {NoStop}%
\bibitem [{\citenamefont {Rotter}\ \emph {et~al.}(2008)\citenamefont {Rotter},
  \citenamefont {Tegel},\ and\ \citenamefont
  {Johrendt}}]{Rotter.M_etal.Phys.-Rev.-Lett.2008}%
  \BibitemOpen
  \bibfield  {author} {\bibinfo {author} {\bibfnamefont {M.}~\bibnamefont
  {Rotter}}, \bibinfo {author} {\bibfnamefont {M.}~\bibnamefont {Tegel}}, \
  and\ \bibinfo {author} {\bibfnamefont {D.}~\bibnamefont {Johrendt}},\ }\Doi
  {10.1103/PhysRevLett.101.107006} {\bibfield  {journal} {\bibinfo  {journal}
  {Phys. Rev. Lett.},\ }\textbf {\bibinfo {volume} {101}},\ \bibinfo {pages}
  {107006} (\bibinfo {year} {2008})}\BibitemShut {NoStop}%
\bibitem [{\citenamefont {Sefat}\ \emph {et~al.}(2008)\citenamefont {Sefat},
  \citenamefont {Jin}, \citenamefont {McGuire}, \citenamefont {Sales},
  \citenamefont {Singh},\ and\ \citenamefont
  {Mandrus}}]{Sefat.A_etal.Phys.-Rev.-Lett.2008}%
  \BibitemOpen
  \bibfield  {author} {\bibinfo {author} {\bibfnamefont {A.~S.}\ \bibnamefont
  {Sefat}}, \bibinfo {author} {\bibfnamefont {R.}~\bibnamefont {Jin}}, \bibinfo
  {author} {\bibfnamefont {M.~A.}\ \bibnamefont {McGuire}}, \bibinfo {author}
  {\bibfnamefont {B.~C.}\ \bibnamefont {Sales}}, \bibinfo {author}
  {\bibfnamefont {D.~J.}\ \bibnamefont {Singh}}, \ and\ \bibinfo {author}
  {\bibfnamefont {D.}~\bibnamefont {Mandrus}},\ }\Doi
  {10.1103/PhysRevLett.101.117004} {\bibfield  {journal} {\bibinfo  {journal}
  {Phys. Rev. Lett.},\ }\textbf {\bibinfo {volume} {101}},\ \bibinfo {pages}
  {117004} (\bibinfo {year} {2008})}\BibitemShut {NoStop}%
\bibitem [{\citenamefont {Ni}\ \emph {et~al.}(2010)\citenamefont {Ni},
  \citenamefont {Thaler}, \citenamefont {Yan}, \citenamefont {Kracher},
  \citenamefont {Colombier}, \citenamefont {Bud'ko}, \citenamefont {Canfield},\
  and\ \citenamefont {Hannahs}}]{Ni.N_etal.Phys.-Rev.-B2010}%
  \BibitemOpen
  \bibfield  {author} {\bibinfo {author} {\bibfnamefont {N.}~\bibnamefont
  {Ni}}, \bibinfo {author} {\bibfnamefont {A.}~\bibnamefont {Thaler}}, \bibinfo
  {author} {\bibfnamefont {J.~Q.}\ \bibnamefont {Yan}}, \bibinfo {author}
  {\bibfnamefont {A.}~\bibnamefont {Kracher}}, \bibinfo {author} {\bibfnamefont
  {E.}~\bibnamefont {Colombier}}, \bibinfo {author} {\bibfnamefont {S.~L.}\
  \bibnamefont {Bud'ko}}, \bibinfo {author} {\bibfnamefont {P.~C.}\
  \bibnamefont {Canfield}}, \ and\ \bibinfo {author} {\bibfnamefont {S.~T.}\
  \bibnamefont {Hannahs}},\ }\Doi {10.1103/PhysRevB.82.024519} {\bibfield
  {journal} {\bibinfo  {journal} {Phys. Rev. B},\ }\textbf {\bibinfo {volume}
  {82}},\ \bibinfo {pages} {024519} (\bibinfo {year} {2010})}\BibitemShut
  {NoStop}%
\bibitem [{\citenamefont {Kim}\ \emph {et~al.}(2010)\citenamefont {Kim},
  \citenamefont {Kreyssig}, \citenamefont {Thaler}, \citenamefont {Pratt},
  \citenamefont {Tian}, \citenamefont {Zarestky}, \citenamefont {Green},
  \citenamefont {Bud'ko}, \citenamefont {Canfield}, \citenamefont {McQueeney},\
  and\ \citenamefont {Goldman}}]{Kim.M_etal.Phys.-Rev.-B2010}%
  \BibitemOpen
  \bibfield  {author} {\bibinfo {author} {\bibfnamefont {M.~G.}\ \bibnamefont
  {Kim}}, \bibinfo {author} {\bibfnamefont {A.}~\bibnamefont {Kreyssig}},
  \bibinfo {author} {\bibfnamefont {A.}~\bibnamefont {Thaler}}, \bibinfo
  {author} {\bibfnamefont {D.~K.}\ \bibnamefont {Pratt}}, \bibinfo {author}
  {\bibfnamefont {W.}~\bibnamefont {Tian}}, \bibinfo {author} {\bibfnamefont
  {J.~L.}\ \bibnamefont {Zarestky}}, \bibinfo {author} {\bibfnamefont {M.~A.}\
  \bibnamefont {Green}}, \bibinfo {author} {\bibfnamefont {S.~L.}\ \bibnamefont
  {Bud'ko}}, \bibinfo {author} {\bibfnamefont {P.~C.}\ \bibnamefont
  {Canfield}}, \bibinfo {author} {\bibfnamefont {R.~J.}\ \bibnamefont
  {McQueeney}}, \ and\ \bibinfo {author} {\bibfnamefont {A.~I.}\ \bibnamefont
  {Goldman}},\ }\Doi {10.1103/PhysRevB.82.220503} {\bibfield  {journal}
  {\bibinfo  {journal} {Phys. Rev. B},\ }\textbf {\bibinfo {volume} {82}},\
  \bibinfo {pages} {220503} (\bibinfo {year} {2010})}\BibitemShut {NoStop}%
\bibitem [{\citenamefont {Thaler}\ \emph {et~al.}(2011)\citenamefont {Thaler},
  \citenamefont {Hodovanets}, \citenamefont {Torikachvili}, \citenamefont
  {Ran}, \citenamefont {Kracher}, \citenamefont {Straszheim}, \citenamefont
  {Yan}, \citenamefont {Mun},\ and\ \citenamefont
  {Canfield}}]{Thaler.A_etal.Phys.-Rev.-B2011}%
  \BibitemOpen
  \bibfield  {author} {\bibinfo {author} {\bibfnamefont {A.}~\bibnamefont
  {Thaler}}, \bibinfo {author} {\bibfnamefont {H.}~\bibnamefont {Hodovanets}},
  \bibinfo {author} {\bibfnamefont {M.~S.}\ \bibnamefont {Torikachvili}},
  \bibinfo {author} {\bibfnamefont {S.}~\bibnamefont {Ran}}, \bibinfo {author}
  {\bibfnamefont {A.}~\bibnamefont {Kracher}}, \bibinfo {author} {\bibfnamefont
  {W.}~\bibnamefont {Straszheim}}, \bibinfo {author} {\bibfnamefont {J.~Q.}\
  \bibnamefont {Yan}}, \bibinfo {author} {\bibfnamefont {E.}~\bibnamefont
  {Mun}}, \ and\ \bibinfo {author} {\bibfnamefont {P.~C.}\ \bibnamefont
  {Canfield}},\ }\Doi {10.1103/PhysRevB.84.144528} {\bibfield  {journal}
  {\bibinfo  {journal} {Phys. Rev. B},\ }\textbf {\bibinfo {volume} {84}},\
  \bibinfo {pages} {144528} (\bibinfo {year} {2011})}\BibitemShut {NoStop}%
\bibitem [{\citenamefont {Ohgushi}\ and\ \citenamefont
  {Kiuchi}(2012)}]{Ohgushi.K_etal.Phys.-Rev.-B2012}%
  \BibitemOpen
  \bibfield  {author} {\bibinfo {author} {\bibfnamefont {K.}~\bibnamefont
  {Ohgushi}}\ and\ \bibinfo {author} {\bibfnamefont {Y.}~\bibnamefont
  {Kiuchi}},\ }\Doi {10.1103/PhysRevB.85.064522} {\bibfield  {journal}
  {\bibinfo  {journal} {Phys. Rev. B},\ }\textbf {\bibinfo {volume} {85}},\
  \bibinfo {pages} {064522} (\bibinfo {year} {2012})}\BibitemShut {NoStop}%
\bibitem [{\citenamefont {Tucker}\ \emph {et~al.}(2012)\citenamefont {Tucker},
  \citenamefont {Pratt}, \citenamefont {Kim}, \citenamefont {Ran},
  \citenamefont {Thaler}, \citenamefont {Granroth}, \citenamefont {Marty},
  \citenamefont {Tian}, \citenamefont {Zarestky}, \citenamefont {Lumsden},
  \citenamefont {Bud'ko}, \citenamefont {Canfield}, \citenamefont {Kreyssig},
  \citenamefont {Goldman},\ and\ \citenamefont
  {McQueeney}}]{Tucker.G_etal.Phys.-Rev.-B2012}%
  \BibitemOpen
  \bibfield  {author} {\bibinfo {author} {\bibfnamefont {G.~S.}\ \bibnamefont
  {Tucker}}, \bibinfo {author} {\bibfnamefont {D.~K.}\ \bibnamefont {Pratt}},
  \bibinfo {author} {\bibfnamefont {M.~G.}\ \bibnamefont {Kim}}, \bibinfo
  {author} {\bibfnamefont {S.}~\bibnamefont {Ran}}, \bibinfo {author}
  {\bibfnamefont {A.}~\bibnamefont {Thaler}}, \bibinfo {author} {\bibfnamefont
  {G.~E.}\ \bibnamefont {Granroth}}, \bibinfo {author} {\bibfnamefont
  {K.}~\bibnamefont {Marty}}, \bibinfo {author} {\bibfnamefont
  {W.}~\bibnamefont {Tian}}, \bibinfo {author} {\bibfnamefont {J.~L.}\
  \bibnamefont {Zarestky}}, \bibinfo {author} {\bibfnamefont {M.~D.}\
  \bibnamefont {Lumsden}}, \bibinfo {author} {\bibfnamefont {S.~L.}\
  \bibnamefont {Bud'ko}}, \bibinfo {author} {\bibfnamefont {P.~C.}\
  \bibnamefont {Canfield}}, \bibinfo {author} {\bibfnamefont {A.}~\bibnamefont
  {Kreyssig}}, \bibinfo {author} {\bibfnamefont {A.~I.}\ \bibnamefont
  {Goldman}}, \ and\ \bibinfo {author} {\bibfnamefont {R.~J.}\ \bibnamefont
  {McQueeney}},\ }\Doi {10.1103/PhysRevB.86.020503} {\bibfield  {journal}
  {\bibinfo  {journal} {Phys. Rev. B},\ }\textbf {\bibinfo {volume} {86}},\
  \bibinfo {pages} {020503} (\bibinfo {year} {2012})}\BibitemShut {NoStop}%
\bibitem [{\citenamefont {Fernandes}\ and\ \citenamefont
  {Millis}(2013)}]{Fernandes.R_etal.Phys.-Rev.-Lett.2013}%
  \BibitemOpen
  \bibfield  {author} {\bibinfo {author} {\bibfnamefont {R.~M.}\ \bibnamefont
  {Fernandes}}\ and\ \bibinfo {author} {\bibfnamefont {A.~J.}\ \bibnamefont
  {Millis}},\ }\Doi {10.1103/PhysRevLett.110.117004} {\bibfield  {journal}
  {\bibinfo  {journal} {Phys. Rev. Lett.},\ }\textbf {\bibinfo {volume}
  {110}},\ \bibinfo {pages} {117004} (\bibinfo {year} {2013})}\BibitemShut
  {NoStop}%
\bibitem [{\citenamefont {Misawa}\ \emph {et~al.}(2012)\citenamefont {Misawa},
  \citenamefont {Nakamura},\ and\ \citenamefont
  {Imada}}]{Misawa.T_etal.Phys.-Rev.-Lett.2012}%
  \BibitemOpen
  \bibfield  {author} {\bibinfo {author} {\bibfnamefont {T.}~\bibnamefont
  {Misawa}}, \bibinfo {author} {\bibfnamefont {K.}~\bibnamefont {Nakamura}}, \
  and\ \bibinfo {author} {\bibfnamefont {M.}~\bibnamefont {Imada}},\ }\Doi
  {10.1103/PhysRevLett.108.177007} {\bibfield  {journal} {\bibinfo  {journal}
  {Phys. Rev. Lett.},\ }\textbf {\bibinfo {volume} {108}},\ \bibinfo {pages}
  {177007} (\bibinfo {year} {2012})}\BibitemShut {NoStop}%
\bibitem [{\citenamefont {Ishida}\ and\ \citenamefont
  {Liebsch}(2010)}]{Ishida.H_etal.Phys.-Rev.-B2010}%
  \BibitemOpen
  \bibfield  {author} {\bibinfo {author} {\bibfnamefont {H.}~\bibnamefont
  {Ishida}}\ and\ \bibinfo {author} {\bibfnamefont {A.}~\bibnamefont
  {Liebsch}},\ }\Doi {10.1103/PhysRevB.81.054513} {\bibfield  {journal}
  {\bibinfo  {journal} {Phys. Rev. B},\ }\textbf {\bibinfo {volume} {81}},\
  \bibinfo {pages} {054513} (\bibinfo {year} {2010})}\BibitemShut {NoStop}%
\bibitem [{\citenamefont {Singh}\ \emph {et~al.}(2009)\citenamefont {Singh},
  \citenamefont {Ellern},\ and\ \citenamefont
  {Johnston}}]{Singh.Y_etal.Phys.-Rev.-B2009}%
  \BibitemOpen
  \bibfield  {author} {\bibinfo {author} {\bibfnamefont {Y.}~\bibnamefont
  {Singh}}, \bibinfo {author} {\bibfnamefont {A.}~\bibnamefont {Ellern}}, \
  and\ \bibinfo {author} {\bibfnamefont {D.~C.}\ \bibnamefont {Johnston}},\
  }\Doi {10.1103/PhysRevB.79.094519} {\bibfield  {journal} {\bibinfo  {journal}
  {Phys. Rev. B},\ }\textbf {\bibinfo {volume} {79}},\ \bibinfo {pages}
  {094519} (\bibinfo {year} {2009})}\BibitemShut {NoStop}%
\bibitem [{\citenamefont {Johnston}\ \emph {et~al.}(2011)\citenamefont
  {Johnston}, \citenamefont {McQueeney}, \citenamefont {Lake}, \citenamefont
  {Honecker}, \citenamefont {Zhitomirsky}, \citenamefont {Nath}, \citenamefont
  {Furukawa}, \citenamefont {Antropov},\ and\ \citenamefont
  {Singh}}]{Johnston.D_etal.Phys.-Rev.-B2011}%
  \BibitemOpen
  \bibfield  {author} {\bibinfo {author} {\bibfnamefont {D.~C.}\ \bibnamefont
  {Johnston}}, \bibinfo {author} {\bibfnamefont {R.~J.}\ \bibnamefont
  {McQueeney}}, \bibinfo {author} {\bibfnamefont {B.}~\bibnamefont {Lake}},
  \bibinfo {author} {\bibfnamefont {A.}~\bibnamefont {Honecker}}, \bibinfo
  {author} {\bibfnamefont {M.~E.}\ \bibnamefont {Zhitomirsky}}, \bibinfo
  {author} {\bibfnamefont {R.}~\bibnamefont {Nath}}, \bibinfo {author}
  {\bibfnamefont {Y.}~\bibnamefont {Furukawa}}, \bibinfo {author}
  {\bibfnamefont {V.~P.}\ \bibnamefont {Antropov}}, \ and\ \bibinfo {author}
  {\bibfnamefont {Y.}~\bibnamefont {Singh}},\ }\Doi
  {10.1103/PhysRevB.84.094445} {\bibfield  {journal} {\bibinfo  {journal}
  {Phys. Rev. B},\ }\textbf {\bibinfo {volume} {84}},\ \bibinfo {pages}
  {094445} (\bibinfo {year} {2011})}\BibitemShut {NoStop}%
\bibitem [{\citenamefont {Nakashima}\ \emph {et~al.}(2010)\citenamefont
  {Nakashima}, \citenamefont {Yui},\ and\ \citenamefont
  {Sasagawa}}]{Nakashima.Y_etal.Physica-C:-Superconductivity2010}%
  \BibitemOpen
  \bibfield  {author} {\bibinfo {author} {\bibfnamefont {Y.}~\bibnamefont
  {Nakashima}}, \bibinfo {author} {\bibfnamefont {H.}~\bibnamefont {Yui}}, \
  and\ \bibinfo {author} {\bibfnamefont {T.}~\bibnamefont {Sasagawa}},\ }\Doi
  {10.1016/j.physc.2010.05.036} {\bibfield  {journal} {\bibinfo  {journal}
  {Physica C},\ }\textbf {\bibinfo {volume} {470}},\ \bibinfo {pages} {1063 }
  (\bibinfo {year} {2010})}\BibitemShut {NoStop}%
\bibitem [{\citenamefont {Regan}\ \emph {et~al.}(2001)\citenamefont {Regan},
  \citenamefont {Ohldag}, \citenamefont {Stamm}, \citenamefont {Nolting},
  \citenamefont {L\"uning}, \citenamefont {St\"ohr},\ and\ \citenamefont
  {White}}]{Regan.T_etal.Phys.-Rev.-B2001}%
  \BibitemOpen
  \bibfield  {author} {\bibinfo {author} {\bibfnamefont {T.~J.}\ \bibnamefont
  {Regan}}, \bibinfo {author} {\bibfnamefont {H.}~\bibnamefont {Ohldag}},
  \bibinfo {author} {\bibfnamefont {C.}~\bibnamefont {Stamm}}, \bibinfo
  {author} {\bibfnamefont {F.}~\bibnamefont {Nolting}}, \bibinfo {author}
  {\bibfnamefont {J.}~\bibnamefont {L\"uning}}, \bibinfo {author}
  {\bibfnamefont {J.}~\bibnamefont {St\"ohr}}, \ and\ \bibinfo {author}
  {\bibfnamefont {R.~L.}\ \bibnamefont {White}},\ }\Doi
  {10.1103/PhysRevB.64.214422} {\bibfield  {journal} {\bibinfo  {journal}
  {Phys. Rev. B},\ }\textbf {\bibinfo {volume} {64}},\ \bibinfo {pages}
  {214422} (\bibinfo {year} {2001})}\BibitemShut {NoStop}%
\bibitem [{\citenamefont {Andrieu}\ \emph {et~al.}(1998)\citenamefont
  {Andrieu}, \citenamefont {Foy}, \citenamefont {Fischer}, \citenamefont
  {Alnot}, \citenamefont {Chevrier}, \citenamefont {Krill},\ and\ \citenamefont
  {Piecuch}}]{Andrieu.S_etal.Phys.-Rev.-B1998}%
  \BibitemOpen
  \bibfield  {author} {\bibinfo {author} {\bibfnamefont {S.}~\bibnamefont
  {Andrieu}}, \bibinfo {author} {\bibfnamefont {E.}~\bibnamefont {Foy}},
  \bibinfo {author} {\bibfnamefont {H.}~\bibnamefont {Fischer}}, \bibinfo
  {author} {\bibfnamefont {M.}~\bibnamefont {Alnot}}, \bibinfo {author}
  {\bibfnamefont {F.}~\bibnamefont {Chevrier}}, \bibinfo {author}
  {\bibfnamefont {G.}~\bibnamefont {Krill}}, \ and\ \bibinfo {author}
  {\bibfnamefont {M.}~\bibnamefont {Piecuch}},\ }\Doi
  {10.1103/PhysRevB.58.8210} {\bibfield  {journal} {\bibinfo  {journal} {Phys.
  Rev. B},\ }\textbf {\bibinfo {volume} {58}},\ \bibinfo {pages} {8210}
  (\bibinfo {year} {1998})}\BibitemShut {NoStop}%
\bibitem [{\citenamefont {Takeda}\ \emph {et~al.}(2008)\citenamefont {Takeda},
  \citenamefont {Kobayashi}, \citenamefont {Okane}, \citenamefont {Ohkochi},
  \citenamefont {Okamoto}, \citenamefont {Saitoh}, \citenamefont {Kobayashi},
  \citenamefont {Yamagami}, \citenamefont {Fujimori}, \citenamefont {Tanaka},
  \citenamefont {Okabayashi}, \citenamefont {Oshima}, \citenamefont {Ohya},
  \citenamefont {Hai},\ and\ \citenamefont
  {Tanaka}}]{Takeda.Y_etal.Phys.-Rev.-Lett.2008}%
  \BibitemOpen
  \bibfield  {author} {\bibinfo {author} {\bibfnamefont {Y.}~\bibnamefont
  {Takeda}}, \bibinfo {author} {\bibfnamefont {M.}~\bibnamefont {Kobayashi}},
  \bibinfo {author} {\bibfnamefont {T.}~\bibnamefont {Okane}}, \bibinfo
  {author} {\bibfnamefont {T.}~\bibnamefont {Ohkochi}}, \bibinfo {author}
  {\bibfnamefont {J.}~\bibnamefont {Okamoto}}, \bibinfo {author} {\bibfnamefont
  {Y.}~\bibnamefont {Saitoh}}, \bibinfo {author} {\bibfnamefont
  {K.}~\bibnamefont {Kobayashi}}, \bibinfo {author} {\bibfnamefont
  {H.}~\bibnamefont {Yamagami}}, \bibinfo {author} {\bibfnamefont
  {A.}~\bibnamefont {Fujimori}}, \bibinfo {author} {\bibfnamefont
  {A.}~\bibnamefont {Tanaka}}, \bibinfo {author} {\bibfnamefont
  {J.}~\bibnamefont {Okabayashi}}, \bibinfo {author} {\bibfnamefont
  {M.}~\bibnamefont {Oshima}}, \bibinfo {author} {\bibfnamefont
  {S.}~\bibnamefont {Ohya}}, \bibinfo {author} {\bibfnamefont {P.~N.}\
  \bibnamefont {Hai}}, \ and\ \bibinfo {author} {\bibfnamefont
  {M.}~\bibnamefont {Tanaka}},\ }\Doi {10.1103/PhysRevLett.100.247202}
  {\bibfield  {journal} {\bibinfo  {journal} {Phys. Rev. Lett.},\ }\textbf
  {\bibinfo {volume} {100}},\ \bibinfo {pages} {247202} (\bibinfo {year}
  {2008})}\BibitemShut {NoStop}%
\bibitem [{\citenamefont {Hwang}\ \emph {et~al.}(2007)\citenamefont {Hwang},
  \citenamefont {Kobayashi}, \citenamefont {Song}, \citenamefont {Fujimori},
  \citenamefont {Tanaka}, \citenamefont {Yang}, \citenamefont {Lin},
  \citenamefont {Huang}, \citenamefont {Chen}, \citenamefont {Jeon},\ and\
  \citenamefont {Kang}}]{Hwang.J_etal.Applied-Physics-Letters2007}%
  \BibitemOpen
  \bibfield  {author} {\bibinfo {author} {\bibfnamefont {J.~I.}\ \bibnamefont
  {Hwang}}, \bibinfo {author} {\bibfnamefont {M.}~\bibnamefont {Kobayashi}},
  \bibinfo {author} {\bibfnamefont {G.~S.}\ \bibnamefont {Song}}, \bibinfo
  {author} {\bibfnamefont {A.}~\bibnamefont {Fujimori}}, \bibinfo {author}
  {\bibfnamefont {A.}~\bibnamefont {Tanaka}}, \bibinfo {author} {\bibfnamefont
  {Z.~S.}\ \bibnamefont {Yang}}, \bibinfo {author} {\bibfnamefont {H.~J.}\
  \bibnamefont {Lin}}, \bibinfo {author} {\bibfnamefont {D.~J.}\ \bibnamefont
  {Huang}}, \bibinfo {author} {\bibfnamefont {C.~T.}\ \bibnamefont {Chen}},
  \bibinfo {author} {\bibfnamefont {H.~C.}\ \bibnamefont {Jeon}}, \ and\
  \bibinfo {author} {\bibfnamefont {T.~W.}\ \bibnamefont {Kang}},\ }\Doi
  {10.1063/1.2769944} {\bibfield  {journal} {\bibinfo  {journal} {Appl. Phys.
  Lett.},\ }\textbf {\bibinfo {volume} {91}},\ \bibinfo {eid} {072507}
  (\bibinfo {year} {2007})}\BibitemShut {NoStop}%
\bibitem [{\citenamefont {Burnus}\ \emph {et~al.}(2008)\citenamefont {Burnus},
  \citenamefont {Hu}, \citenamefont {Hsieh}, \citenamefont {Joly},
  \citenamefont {Joy}, \citenamefont {Haverkort}, \citenamefont {Wu},
  \citenamefont {Tanaka}, \citenamefont {Lin}, \citenamefont {Chen},\ and\
  \citenamefont {Tjeng}}]{Burnus.T_etal.Phys.-Rev.-B2008}%
  \BibitemOpen
  \bibfield  {author} {\bibinfo {author} {\bibfnamefont {T.}~\bibnamefont
  {Burnus}}, \bibinfo {author} {\bibfnamefont {Z.}~\bibnamefont {Hu}}, \bibinfo
  {author} {\bibfnamefont {H.~H.}\ \bibnamefont {Hsieh}}, \bibinfo {author}
  {\bibfnamefont {V.~L.~J.}\ \bibnamefont {Joly}}, \bibinfo {author}
  {\bibfnamefont {P.~A.}\ \bibnamefont {Joy}}, \bibinfo {author} {\bibfnamefont
  {M.~W.}\ \bibnamefont {Haverkort}}, \bibinfo {author} {\bibfnamefont
  {H.}~\bibnamefont {Wu}}, \bibinfo {author} {\bibfnamefont {A.}~\bibnamefont
  {Tanaka}}, \bibinfo {author} {\bibfnamefont {H.-J.}\ \bibnamefont {Lin}},
  \bibinfo {author} {\bibfnamefont {C.~T.}\ \bibnamefont {Chen}}, \ and\
  \bibinfo {author} {\bibfnamefont {L.~H.}\ \bibnamefont {Tjeng}},\ }\Doi
  {10.1103/PhysRevB.77.125124} {\bibfield  {journal} {\bibinfo  {journal}
  {Phys. Rev. B},\ }\textbf {\bibinfo {volume} {77}},\ \bibinfo {pages}
  {125124} (\bibinfo {year} {2008})}\BibitemShut {NoStop}%
\bibitem [{\citenamefont {Rader}\ \emph {et~al.}(2004)\citenamefont {Rader},
  \citenamefont {Pampuch}, \citenamefont {Shikin}, \citenamefont {Gudat},
  \citenamefont {Okabayashi}, \citenamefont {Mizokawa}, \citenamefont
  {Fujimori}, \citenamefont {Hayashi}, \citenamefont {Tanaka}, \citenamefont
  {Tanaka},\ and\ \citenamefont {Kimura}}]{Rader.O_etal.Phys.-Rev.-B2004}%
  \BibitemOpen
  \bibfield  {author} {\bibinfo {author} {\bibfnamefont {O.}~\bibnamefont
  {Rader}}, \bibinfo {author} {\bibfnamefont {C.}~\bibnamefont {Pampuch}},
  \bibinfo {author} {\bibfnamefont {A.~M.}\ \bibnamefont {Shikin}}, \bibinfo
  {author} {\bibfnamefont {W.}~\bibnamefont {Gudat}}, \bibinfo {author}
  {\bibfnamefont {J.}~\bibnamefont {Okabayashi}}, \bibinfo {author}
  {\bibfnamefont {T.}~\bibnamefont {Mizokawa}}, \bibinfo {author}
  {\bibfnamefont {A.}~\bibnamefont {Fujimori}}, \bibinfo {author}
  {\bibfnamefont {T.}~\bibnamefont {Hayashi}}, \bibinfo {author} {\bibfnamefont
  {M.}~\bibnamefont {Tanaka}}, \bibinfo {author} {\bibfnamefont
  {A.}~\bibnamefont {Tanaka}}, \ and\ \bibinfo {author} {\bibfnamefont
  {A.}~\bibnamefont {Kimura}},\ }\Doi {10.1103/PhysRevB.69.075202} {\bibfield
  {journal} {\bibinfo  {journal} {Phys. Rev. B},\ }\textbf {\bibinfo {volume}
  {69}},\ \bibinfo {pages} {075202} (\bibinfo {year} {2004})}\BibitemShut
  {NoStop}%
\bibitem [{\citenamefont {Gel'mukhanov}\ and\ \citenamefont
  {{\AA}gren}(1999)}]{Gelmukhanov.F_etal.Physics-Reports1999}%
  \BibitemOpen
  \bibfield  {author} {\bibinfo {author} {\bibfnamefont {F.}~\bibnamefont
  {Gel'mukhanov}}\ and\ \bibinfo {author} {\bibfnamefont {H.}~\bibnamefont
  {{\AA}gren}},\ }\href@noop {} {\bibfield  {journal} {\bibinfo  {journal}
  {Phys. Rep.},\ }\textbf {\bibinfo {volume} {312}},\ \bibinfo {pages} {87}
  (\bibinfo {year} {1999})}\BibitemShut {NoStop}%
\bibitem [{\citenamefont {Br\"uhwiler}\ \emph {et~al.}(2002)\citenamefont
  {Br\"uhwiler}, \citenamefont {Karis},\ and\ \citenamefont
  {M\aa{}rtensson}}]{Bruhwiler.P_etal.Rev.-Mod.-Phys.2002}%
  \BibitemOpen
  \bibfield  {author} {\bibinfo {author} {\bibfnamefont {P.~A.}\ \bibnamefont
  {Br\"uhwiler}}, \bibinfo {author} {\bibfnamefont {O.}~\bibnamefont {Karis}},
  \ and\ \bibinfo {author} {\bibfnamefont {N.}~\bibnamefont {M\aa{}rtensson}},\
  }\Doi {10.1103/RevModPhys.74.703} {\bibfield  {journal} {\bibinfo  {journal}
  {Rev. Mod. Phys.},\ }\textbf {\bibinfo {volume} {74}},\ \bibinfo {pages}
  {703} (\bibinfo {year} {2002})}\BibitemShut {NoStop}%
\bibitem [{\citenamefont {Levy}\ \emph {et~al.}(2012)\citenamefont {Levy},
  \citenamefont {Sutarto}, \citenamefont {Chevrier}, \citenamefont {Regier},
  \citenamefont {Blyth}, \citenamefont {Geck}, \citenamefont {Wurmehl},
  \citenamefont {Harnagea}, \citenamefont {Wadati}, \citenamefont {Mizokawa},
  \citenamefont {Elfimov}, \citenamefont {Damascelli},\ and\ \citenamefont
  {Sawatzky}}]{Levy.G_etal.Phys.-Rev.-Lett.2012}%
  \BibitemOpen
  \bibfield  {author} {\bibinfo {author} {\bibfnamefont {G.}~\bibnamefont
  {Levy}}, \bibinfo {author} {\bibfnamefont {R.}~\bibnamefont {Sutarto}},
  \bibinfo {author} {\bibfnamefont {D.}~\bibnamefont {Chevrier}}, \bibinfo
  {author} {\bibfnamefont {T.}~\bibnamefont {Regier}}, \bibinfo {author}
  {\bibfnamefont {R.}~\bibnamefont {Blyth}}, \bibinfo {author} {\bibfnamefont
  {J.}~\bibnamefont {Geck}}, \bibinfo {author} {\bibfnamefont {S.}~\bibnamefont
  {Wurmehl}}, \bibinfo {author} {\bibfnamefont {L.}~\bibnamefont {Harnagea}},
  \bibinfo {author} {\bibfnamefont {H.}~\bibnamefont {Wadati}}, \bibinfo
  {author} {\bibfnamefont {T.}~\bibnamefont {Mizokawa}}, \bibinfo {author}
  {\bibfnamefont {I.~S.}\ \bibnamefont {Elfimov}}, \bibinfo {author}
  {\bibfnamefont {A.}~\bibnamefont {Damascelli}}, \ and\ \bibinfo {author}
  {\bibfnamefont {G.~A.}\ \bibnamefont {Sawatzky}},\ }\Doi
  {10.1103/PhysRevLett.109.077001} {\bibfield  {journal} {\bibinfo  {journal}
  {Phys. Rev. Lett.},\ }\textbf {\bibinfo {volume} {109}},\ \bibinfo {pages}
  {077001} (\bibinfo {year} {2012})}\BibitemShut {NoStop}%
\bibitem [{\citenamefont {Ideta}\ \emph
  {et~al.}(2013){\natexlab{a}}\citenamefont {Ideta}, \citenamefont {Yoshida},
  \citenamefont {Nishi}, \citenamefont {Fujimori}, \citenamefont {Kotani},
  \citenamefont {Ono}, \citenamefont {Nakashima}, \citenamefont {Yamaichi},
  \citenamefont {Sasagawa}, \citenamefont {Nakajima}, \citenamefont {Kihou},
  \citenamefont {Tomioka}, \citenamefont {Lee}, \citenamefont {Iyo},
  \citenamefont {Eisaki}, \citenamefont {Ito}, \citenamefont {Uchida},\ and\
  \citenamefont {Arita}}]{Ideta.S_etal.Phys.-Rev.-Lett.2013}%
  \BibitemOpen
  \bibfield  {author} {\bibinfo {author} {\bibfnamefont {S.}~\bibnamefont
  {Ideta}}, \bibinfo {author} {\bibfnamefont {T.}~\bibnamefont {Yoshida}},
  \bibinfo {author} {\bibfnamefont {I.}~\bibnamefont {Nishi}}, \bibinfo
  {author} {\bibfnamefont {A.}~\bibnamefont {Fujimori}}, \bibinfo {author}
  {\bibfnamefont {Y.}~\bibnamefont {Kotani}}, \bibinfo {author} {\bibfnamefont
  {K.}~\bibnamefont {Ono}}, \bibinfo {author} {\bibfnamefont {Y.}~\bibnamefont
  {Nakashima}}, \bibinfo {author} {\bibfnamefont {S.}~\bibnamefont {Yamaichi}},
  \bibinfo {author} {\bibfnamefont {T.}~\bibnamefont {Sasagawa}}, \bibinfo
  {author} {\bibfnamefont {M.}~\bibnamefont {Nakajima}}, \bibinfo {author}
  {\bibfnamefont {K.}~\bibnamefont {Kihou}}, \bibinfo {author} {\bibfnamefont
  {Y.}~\bibnamefont {Tomioka}}, \bibinfo {author} {\bibfnamefont {C.~H.}\
  \bibnamefont {Lee}}, \bibinfo {author} {\bibfnamefont {A.}~\bibnamefont
  {Iyo}}, \bibinfo {author} {\bibfnamefont {H.}~\bibnamefont {Eisaki}},
  \bibinfo {author} {\bibfnamefont {T.}~\bibnamefont {Ito}}, \bibinfo {author}
  {\bibfnamefont {S.}~\bibnamefont {Uchida}}, \ and\ \bibinfo {author}
  {\bibfnamefont {R.}~\bibnamefont {Arita}},\ }\Doi
  {10.1103/PhysRevLett.110.107007} {\bibfield  {journal} {\bibinfo  {journal}
  {Phys. Rev. Lett.},\ }\textbf {\bibinfo {volume} {110}},\ \bibinfo {pages}
  {107007} (\bibinfo {year} {2013}{\natexlab{a}})}\BibitemShut {NoStop}%
\bibitem [{\citenamefont {Konbu}\ \emph {et~al.}(2012)\citenamefont {Konbu},
  \citenamefont {Nakamura}, \citenamefont {Ikeda},\ and\ \citenamefont
  {Arita}}]{Konbu.S_etal.Solid-State-Communications2012}%
  \BibitemOpen
  \bibfield  {author} {\bibinfo {author} {\bibfnamefont {S.}~\bibnamefont
  {Konbu}}, \bibinfo {author} {\bibfnamefont {K.}~\bibnamefont {Nakamura}},
  \bibinfo {author} {\bibfnamefont {H.}~\bibnamefont {Ikeda}}, \ and\ \bibinfo
  {author} {\bibfnamefont {R.}~\bibnamefont {Arita}},\ }\Doi
  {10.1016/j.ssc.2011.12.048} {\bibfield  {journal} {\bibinfo  {journal} {Solid
  State Commun.},\ }\textbf {\bibinfo {volume} {152}},\ \bibinfo {pages} {728 }
  (\bibinfo {year} {2012})}\BibitemShut {NoStop}%
\bibitem [{\citenamefont {Ideta}\ \emph
  {et~al.}(2013){\natexlab{b}}\citenamefont {Ideta}, \citenamefont {Yoshida},
  \citenamefont {Nakajima}, \citenamefont {Malaeb}, \citenamefont {Shimojima},
  \citenamefont {Ishizaka}, \citenamefont {Fujimori}, \citenamefont
  {Kimigashira}, \citenamefont {Ono}, \citenamefont {Kihou}, \citenamefont
  {Tomioka}, \citenamefont {Lee}, \citenamefont {Iyo}, \citenamefont {Eisaki},
  \citenamefont {Ito},\ and\ \citenamefont
  {Uchida}}]{Ideta.S_etal.Phys.-Rev.-B2013}%
  \BibitemOpen
  \bibfield  {author} {\bibinfo {author} {\bibfnamefont {S.}~\bibnamefont
  {Ideta}}, \bibinfo {author} {\bibfnamefont {T.}~\bibnamefont {Yoshida}},
  \bibinfo {author} {\bibfnamefont {M.}~\bibnamefont {Nakajima}}, \bibinfo
  {author} {\bibfnamefont {W.}~\bibnamefont {Malaeb}}, \bibinfo {author}
  {\bibfnamefont {T.}~\bibnamefont {Shimojima}}, \bibinfo {author}
  {\bibfnamefont {K.}~\bibnamefont {Ishizaka}}, \bibinfo {author}
  {\bibfnamefont {A.}~\bibnamefont {Fujimori}}, \bibinfo {author}
  {\bibfnamefont {H.}~\bibnamefont {Kimigashira}}, \bibinfo {author}
  {\bibfnamefont {K.}~\bibnamefont {Ono}}, \bibinfo {author} {\bibfnamefont
  {K.}~\bibnamefont {Kihou}}, \bibinfo {author} {\bibfnamefont
  {Y.}~\bibnamefont {Tomioka}}, \bibinfo {author} {\bibfnamefont {C.~H.}\
  \bibnamefont {Lee}}, \bibinfo {author} {\bibfnamefont {A.}~\bibnamefont
  {Iyo}}, \bibinfo {author} {\bibfnamefont {H.}~\bibnamefont {Eisaki}},
  \bibinfo {author} {\bibfnamefont {T.}~\bibnamefont {Ito}}, \ and\ \bibinfo
  {author} {\bibfnamefont {S.}~\bibnamefont {Uchida}},\ }\Doi
  {10.1103/PhysRevB.87.201110} {\bibfield  {journal} {\bibinfo  {journal}
  {Phys. Rev. B},\ }\textbf {\bibinfo {volume} {87}},\ \bibinfo {pages}
  {201110} (\bibinfo {year} {2013}{\natexlab{b}})}\BibitemShut {NoStop}%
\bibitem [{\citenamefont {Urata}\ \emph {et~al.}()\citenamefont {Urata},
  \citenamefont {Tanabe}, \citenamefont {Huynh}, \citenamefont {Oguro},
  \citenamefont {Watanabe}, \citenamefont {Heguri},\ and\ \citenamefont
  {Tanigaki}}]{Urata.T_etal.ArXiv-eprints2013}%
  \BibitemOpen
  \bibfield  {author} {\bibinfo {author} {\bibfnamefont {T.}~\bibnamefont
  {Urata}}, \bibinfo {author} {\bibfnamefont {Y.}~\bibnamefont {Tanabe}},
  \bibinfo {author} {\bibfnamefont {K.~K.}\ \bibnamefont {Huynh}}, \bibinfo
  {author} {\bibfnamefont {H.}~\bibnamefont {Oguro}}, \bibinfo {author}
  {\bibfnamefont {K.}~\bibnamefont {Watanabe}}, \bibinfo {author}
  {\bibfnamefont {S.}~\bibnamefont {Heguri}}, \ and\ \bibinfo {author}
  {\bibfnamefont {K.}~\bibnamefont {Tanigaki}},\ }\href@noop {} {}\Eprint
  {http://arxiv.org/abs/1307.2813} {arXiv:1307.2813} \BibitemShut {NoStop}%
\bibitem [{\citenamefont {Georges}\ \emph {et~al.}(2013)\citenamefont
  {Georges}, \citenamefont {Medici},\ and\ \citenamefont
  {Mravlje}}]{Georges.A_etal.Annual-Review-of-Condensed-Matter-Physics2013}%
  \BibitemOpen
  \bibfield  {author} {\bibinfo {author} {\bibfnamefont {A.}~\bibnamefont
  {Georges}}, \bibinfo {author} {\bibfnamefont {L.~d.}\ \bibnamefont {Medici}},
  \ and\ \bibinfo {author} {\bibfnamefont {J.}~\bibnamefont {Mravlje}},\ }\Doi
  {10.1146/annurev-conmatphys-020911-125045} {\bibfield  {journal} {\bibinfo
  {journal} {Annu. Rev. Cond. Mat.},\ }\textbf {\bibinfo {volume} {4}},\
  \bibinfo {pages} {137} (\bibinfo {year} {2013})}\BibitemShut {NoStop}%
\end{thebibliography}%

\end{document}